\documentstyle[preprint,eqsecnum,aps]{revtex}

\begin{document}
\draft
\preprint{}
\title{Observation of Magnetic Moments in the Superconducting State of YBa$_2$Cu$_3$O$_{6.6}$\\}
\author{H. A. Mook, $^1$ Pengcheng Dai,$^1$ F. Do$\rm\breve{g}$an,$^2$}
\address{$^1$Solid State Division, Oak Ridge National Laboratory, Oak Ridge, Tennessee 37831-6393}
\address{$^2$Department of Materials Science and Engineering, University of Washington, Seattle, Washington 98195}
\date{\today}
\maketitle
\begin{abstract}
Neutron Scattering measurements for YBa$_2$Cu$_3$O$_{6.6}$ have identified small magnetic moments that increase in strength as the temperature is reduced below $T^\ast$ and further increase below $T_c$. An analysis of the data shows the moments are antiferromagnetic between the Cu-O planes with a correlation length of longer than 195 \AA\ in the $a$-$b$ plane and about 35 \AA\ along the $c$-axis. The origin of the moments is unknown, and their properties are discusssed 
both in terms of Cu spin magnetism and orbital bond currents.
\end{abstract}

\narrowtext
Since the parent compounds of the high-$T_c$ cuprate superconductors have antiferromagnetic 
\cite{1,2} order the role of magnetism in the superconducting process has been widely investigated. Neutron scattering has played a central role in these investigations with many of the measurements being made on the YBa$_2$Cu$_3$O$_{6+x}$ family of materials. Considerable prior work has been accomplished on the 
YBa$_2$Cu$_3$O$_{6.6}$ (YBCO6.6, $T_c=62.7$ K) crystal utilized in the present study and these have been reviewed in a recent paper by Dai {\it et al.} \cite{3}. Both the resonance \cite{4,5} at 34 meV and 
the incommensurate \cite{6,7,3} spin fluctuations above and below the resonance have been studied extensively. In this paper we present results for a new magnetic scattering feature not observed previously in this material. The search for the magnetic scattering was made in response to a model of the pseudogap proposed by Chakravarty {\it et al.} \cite{8} that invokes an order-parameter competition to explain the temperature dependence of $T_c$ as doping is increased. The order-parameter chosen consists of orbital antiferromagnetism developed by bond currents \cite{8a} in the Cu-O planes. These currents, if they are sufficiently strong, should produce a signal observable by neutron scattering.

Searches for such a signal have resulted in observations that may be consistent with the Chakravarty {\it et al.} \cite{8} proposal. The reciprocal lattice positions needed to check the proposal are the same ones where antiferromagnetism of the Cu spins is observed. Thus it would be expected that the most likely origin of the observed signals would be from Cu spin magnetism. However, the newly found scattering occurs in the superconducting state at zero energy transfer. At low temperatures a spin gap of about 20 meV occurs in the fluctuation spectra \cite{5} and we have previously observed no excitations below this gap. In addition the behavior of the magnetism in the present experiment is quite different than that found in earlier investigations.

The experiment was performed at the HB1 triple-axis spectrometer at the HFIR reactor at Oak Ridge National Laboratory. Pyrolitic graphite monochromator and analyzer crystals were utilized in the experiment with a collimation of 48-40-40-70 minutes from in front of the monochromator to after the analyzer crystal. The neutron energy utilized was 13.78 meV and four pyrolitic graphite filters each about 2 inches thick were placed in the beam to eliminate higher order contamination. The crystal was mounted in the (h, h, l) zone and the (1, 1, 0) reflection gave a counting rate of about 79,000 counts/sec when corrected for neutron absorbers placed in the beam for the crystal alignment.

The crystal was cooled to 10 K using a displex refrigerator and after alignment, measurements at the (0.5, 0.5, l) positions were performed. Measurements at (0.5, 0.5, 0) position showed no peak thus determining the antiferromagnetic nature of the scattering between the Cu-O layers as discussed by Tranquada {\it et al.} \cite{2}.  The scans through the (0.5, 0.5, 0) were used to determine the background scattering. It largely results from the spin and isotopic incoherent cross-sections of the elements in the YBCO structure. The featureless scan at (0.5, 0.5, 0) had an average value of 1830 counts per 5 minutes at 10 K. This background value will be subtracted uniformly from the 10 K data shown in the figures to follow. Results of a measurement for the (0.5, 0.5, 2) reflection are shown in Fig. 1a. For each figure a number of scans were averaged to obtain the data with the counting errors shown. A Gaussian  fit to the data yields a height of $263\pm15$ counts and a width of $0.016\pm0.002$ reciprocal lattice units (r.l.u.) FWHM along the scan direction. A measurement of this reflection at 250 K is shown in Fig.1b. The background determined at (0.5, 0.5, 0) was found to be 30 counts lower than at 10 K.  A scan through the (0.5, 0.5, 1) reflection is shown in Fig 1c. The fit yields a height of $307\pm25$ counts and a width of $0.015\pm0.002$ r.l.u. The measured spectrometer resolution for the scan direction used is 0.014 r.l.u so that the scans are resolution limited within the errors.

We don't know the origin of the moments that produce the scattering. However, a logical way to proceed is to analyze the data in terms of Cu spin magnetism and see if this can account for the results. We can neglect spectrometer resolution effects in taking the ratio of the intensities for the (0.5, 0.5, 1) and (0.5, 0.5, 2) reflections so this is just given by the ratio of the peak heights giving $1.16\pm 0.16$.  Assuming a collinear structure the ratio of intensities between the (0.5, 0.5, 1) and the (0.5, 0.5, 2) are calculated to be 0.98 if the moment is along the 
$c$-axis and 0.52 if the moment is in the $a$-$b$ plane. The data are consistent with the moment pointing 
along the $c$-axis.

A scan along the $c^\ast$ direction is shown in Fig. 1d for the (0.5, 0.5, 1) reflection. 
The peak width along $c^\ast$ is much wider than the resolution and a Gaussian fit yields a 
height of $310\pm20$ counts and a width of $0.343\pm0.025$ r.l.u along $c^\ast$. 
If we define a correlation length as $2\pi/$HWHM(\AA$^{-1}$), we obtain a correlation length of about 195 \AA\ in the $(a,b)$ plane and 35 \AA\ along the $c$ direction. The in-plane scans are resolution limited so 195 \AA\ is a lower limit. Fig. 2a shows a scan at the (0.5, 0.5, 1.1) position. Since as shown in Fig. 1d the scattering is broad along $c^\ast$ we find as expected the intensity to be the same as the scan at (0.5, 0.5, 1).  Unfortunately the crystal contains extraneous phases, particularly Y$_2$BaCuO$_5$, in random orientations. Powder diffraction lines from these phases interfere with the magnetic measurements especially at wavevectors larger than those for the (0.5, 0.5, 2) reflection. A large powder peak interferes with the (0.5, 0.5, 5) reflection, however, a measurement at (0.5, 0.5, 5.1) results in manageable sloping background as shown in Fig. 2b. If the moment was in the $(a,b)$ plane a peak would be expected at this position that is larger in area than at (0.5, 0.5, 1) assuming a Cu form factor. The spectrometer resolution would broaden the peak at (0.5, 0.5, 5.1) by about a factor of 1.4 compared to the (0.5, 0.5, l) , however, 
 an $(a,b)$ moment direction seems to be ruled out. For a $c$-axis moment a peak would be expected that is 4.4  times smaller in area than the (0.5, 0.5, 1) reflection. It appears that even a peak this big may not be visible although it is hard to be definite given the quality of the data. Should this be the case it would mean the scattering falls off more quickly than the Cu form factor ruling out Cu spin magnetism as the source of the scattering.

The scattering at the (1.5, 1.5, 1) and (1.5, 1.5, 2) positions are badly contaminated by extraneous scattering and no relief is found by going off slightly along $c^\ast$ as we did for the (0.5, 0.5, 5) reflection. Fig. 2c shows the scan through the (1.5, 1.5, 1) position. With unpolarized neutrons the best way to obtain the intensity for the (1.5, 1.5, l) and (1.5, 1.5, 2) reflections is to take the difference in the scattering between 10K and 250K. This increases greatly the counting errors so that we have combined our measurements for the (1.5, 1.5, 1) and (1.5, 1.5, 2) reflections. The difference in scattering between 10 K and 250 K is shown in Fig. 2d.
 For a $c$-axis moment a peak 0.3 the area of the (0.5, 0.5, 1)is expected in the temperature difference using the Cu form factor. 
Spectrometer resolution would broaden the peak by a factor of 1.5, but it appears that the measurements show the scattering falls off faster than the Cu form factor. However, we are taking the difference between rather widely separated temperatures and the sharply sloping background makes the interpretation of the results difficult. 
In summary, if Cu spin magnetism is responsible for the observed 
moments, the moment direction appears to be along the $c$-axis. However, it is unclear that Cu spin magnetism 
is responsible and the lack of any observed signal at higher order reflections throw suspicion on 
this interpretation. Cleaner determinations of the higher order reflections are needed to resolve this.

The temperature dependence of the scattering measured at the (0.5, 0.5, 2) position is shown in Fig.3. 
A number of measurements were averaged to obtain the data with the errors shown.
A noticeable increase in intensity is found below $T_c$. The intensity was measured cooling and warming from 10 K to 100 K with repeatable results. The intensity drops off rapidly above 160 K which is in the neighborhood of the pseudogap \cite{9} temperature $T^\ast$. Intensity above the (0.5, 0.5, 0) background at 250 K, which is the zero used in Fig 3, is noticeable up to room temperature. The 30 counts per 5 minutes change in background between 10 K and 250 K mentioned above affects the plot of the temperature dependence to some degree. Since we felt the change across $T_c$ to be important, we monitored the background from 10 K to 120 K and found it unchanged, so that a background shift cannot be responsible for the change at $T_c$.

The observed magnetic scattering is very small and it is difficult to make intensity comparisons to the nuclear Bragg scattering since these reflections are greatly affected by extinction effects. If we assume that the relatively small (0, 0, 2) peak is extinction free and use it to calculate the size of the moment we find that observed scattering is about 2500 times smaller than that expected for an ordered Cu moment of 1$\mu_B$. This number takes account of the observation that the scattering is spread out along $c^\ast$ relative to the spectrometer resolution. Assuming the Cu form factor this would imply a moment about 50 times smaller, since the scattering varies as the moment squared. We have done energy scans at the (0.5, 0.5, 1) and (0.5, 0.5, 0) positions and they both look identical giving no sign that the scattering is inelastic or quasielastic. The energy resolution used was about 1 meV. It is possible that the moments do fluctuate, but on a sufficiently long time scale that the fluctuations are not observed with the energy resolution used.

Neidermayer {\it et al.} \cite{10} find using muon spin rotation that antiferromagnetism in a spin-glass-like state extends into the superconducting region for the cuprate materials. The hole doping for our YBCO6.6 crystal is found to be 0.1 using the relationship of doping to $T_c$ developed by Tallon {\it et al.} 
\cite{11}. This is about the hole doping at which the muon signal ceases to be observed. Furthermore the spin glass magnetism, which can be easily observed with neutrons at lower doping manifests itself in a different pattern. Fig 4 shows measurements for a YBCO6.35 crystal. The observed scattering for the (0.5, 0.5, 5) reflection is much larger than the signal for YBCO6.6 and displays a broad lorentzian distribution as expected for a spin-glass. YBCO6.35 is expected to have a hole doping of about 0.05 for which the spin-glass freezing temperature in the Y$_{1-x}$Ca$_x$Ba$_2$Cu$_3$O$_{6.02}$ system is found 
to be about 20 K for a similar hole doping. Assuming the two materials have similar properties the temperature dependence of the scattering shown in Fig. 4a and b is roughly what might be expected. The ratio of the intensity of the scattering at the (0.5, 0.5, 5) position to that at the (0.5, 0.5, 2) position is found to be $0.72\pm0.1$ showing that the moment is in the $a$-$b$ plane in this case. To produce the present results the spin-glass magnetism would have to rotate to the c-axis with increased doping, and display a rather different temperature dependence. Also the scattering can not be from islands of the insulating phase as these would have to extend more than 195 \AA\ in the $(a,b)$ plane and they also have the wrong moment direction. In addition the temperature dependence of the scattering from such islands would not be expected to show the small jump that appears to occur at $T_c$.

Hsu {\it et al.} \cite{8a} have made predictions for the neutron scattering cross section from orbital currents. Given our (0.5, 0.5, 1) peak of 307 counts, their prediction for the (0.5, 0.5, 2) and (0.5, 0.5, 5) peaks would be 190 and 36 counts respectively, neglecting resolution effects. The value for (0.5, 0.5, 2) is somewhat lower then we would expect, but the (0.5, 0.5, 5) peak is sufficiently small to agree with our measurement. However, they assume a $c$-axis moment and it is not clear that this is appropriate given opposing orbital currents in the bilayers, as the neutron wavelength is 
comparable in size to the current path lengths. 

The present experiment appears to be consistent with the orbital magnetism picture, as the observed moments have the correct size, and temperature dependence. The next experimental step would be to attempt polarized beam measurements to clearly determine the size of the higher order reflections so that the form factor could be used to differentiate between spin and orbital bond currents. On the theoretical side, a calculation for the cross section for orbital bond currents is needed that properly takes into account the size of the bond current paths relative to the neutron wavelength.

In any case, it is surprising to find magnetic moments with unusual properties deep within the superconducting state. The experiment adds to a rich variety of behavior that magnetism has displayed in the cuprate superconductors. The present results are just the first step in determining how the newly observed magnetism fits in with the orbital moment picture. 

The authors appreciate helpful interactions with S. Chakavarty, R. B. Laughlin, Patrick Lee, and Dirk Morr. The structure factor calculations were performed by B. Chakoumakos. 
This work was supported by U.S. DOE under contract 
DE-AC05-00OR22725 with 
UT-Battelle, LLC.

\begin{figure}
\caption{a) Scan along h, h, through the (0.5, 0.5, 2) reflection at 10 K. b) shows the same scan at 250 K. c) Scan along h, h, for the (0.5, 0.5, 1) reflection at 10 K. d) Scan along l for the (0.5, 0.5, 1) reflection at 10 K. The lines are gaussian fits to the data. A background determined from the scans at (0.5, 0.5, 0) has been subtracted from the data. The counting time was 5 min per point with multiple scans averaged to obtain the counting errors shown.
}
\label{autonum}
\end{figure}

\begin{figure}
\caption{
a) Scan along h, h, for the (0.5, 0.5, 1.1) position at 10 K. The line is a gaussian fit to the data. b) Scan along h, h, for the (0.5, 0.5, 5.1) position. The sloping background stems from the tail of a large extraneous powder peak. The line is a linear fit through the data points.  c) Scan along h, h, for the (1.5, 1.5, 1) reflection at 10 K. The scattering is badly contaminated by an extraneous powder peak. d) difference of the scattering at (h, h, 1) summed with that at (h, h, 2) between 10 K and 250 K. The line is a linear fit to the data.  
}
\end{figure}

\begin{figure}
\caption{
Temperature dependence of the scattering at the (0.5, 0.5, 2) position. The background was monitored between 10 K and 120 K and found to be essentially unchanged. The background found at 250 K has been subtracted from the data. Multiple measurements were averaged to obtain the counting errors shown. The line is a weighted fit through the data points. 
}
\end{figure}

\begin{figure}
\caption{
a) scan along h, h, for the (0.5, 0.5, 5) reflection at 10 K for YBCO6.35. Note that the scale is considerably expanded relative to Fig.1a. b) Same scan as a) at 60 K. Backgrounds have been subtracted as in Fig. 1. The lines are lorentzian fits to the data points.         
}
\end{figure}
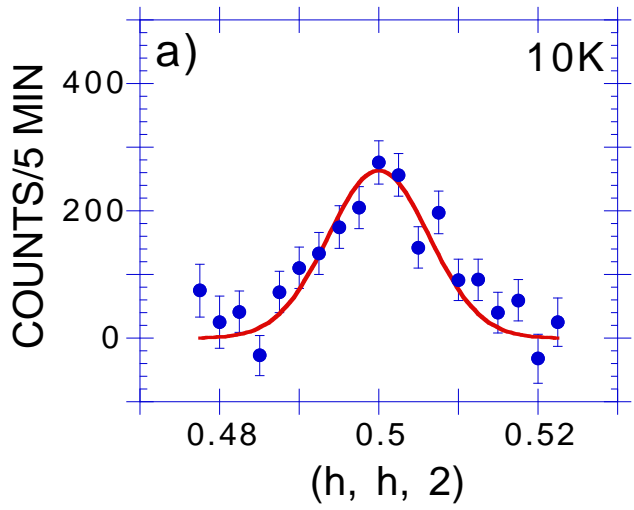
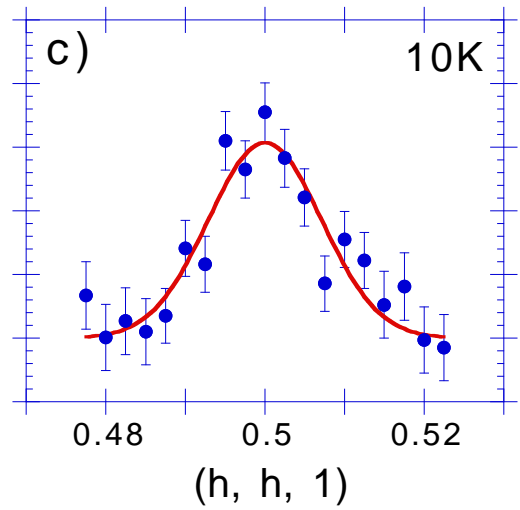
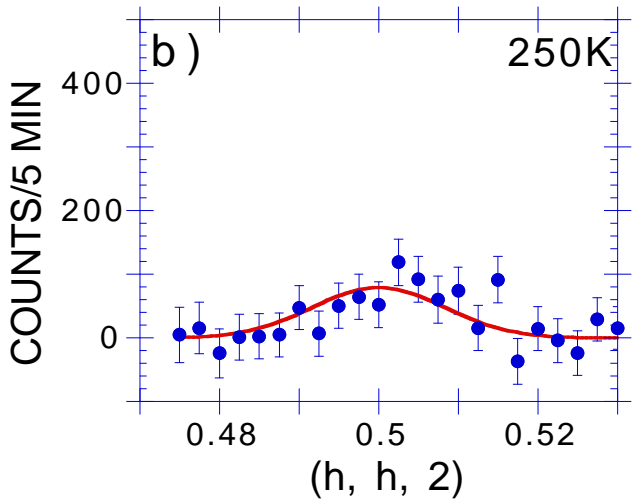
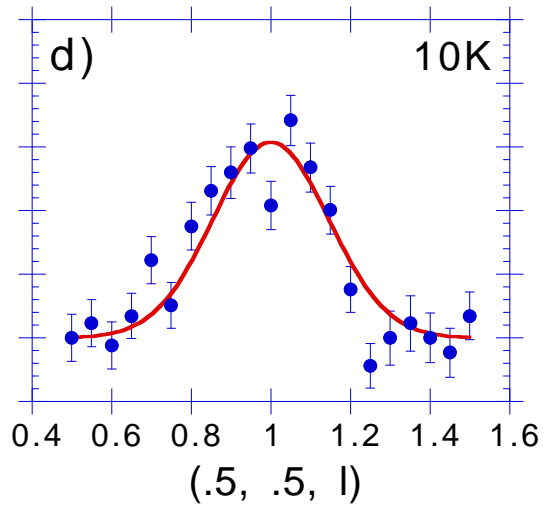
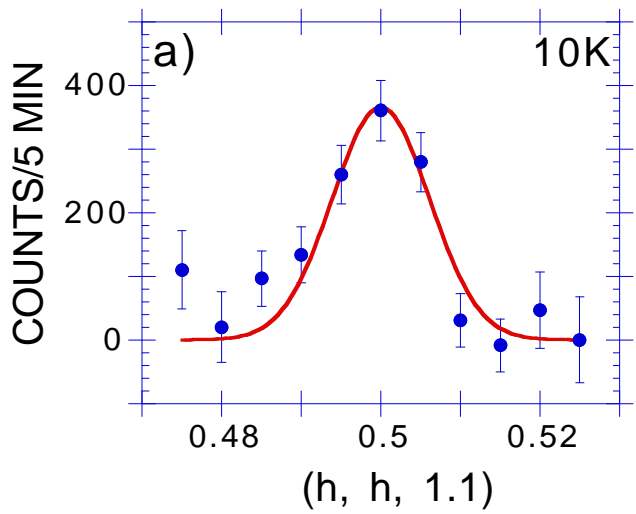
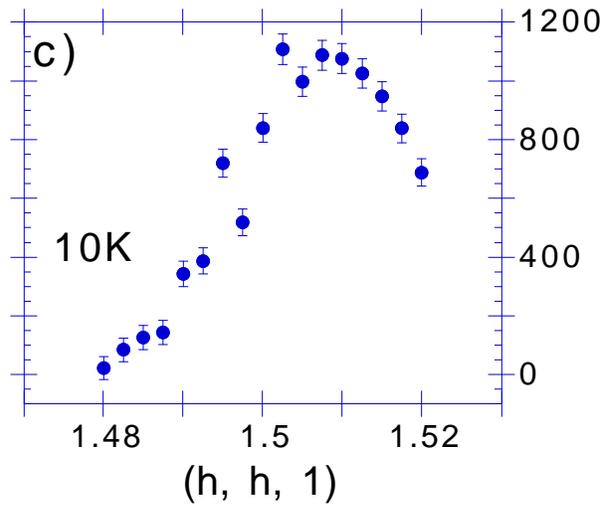
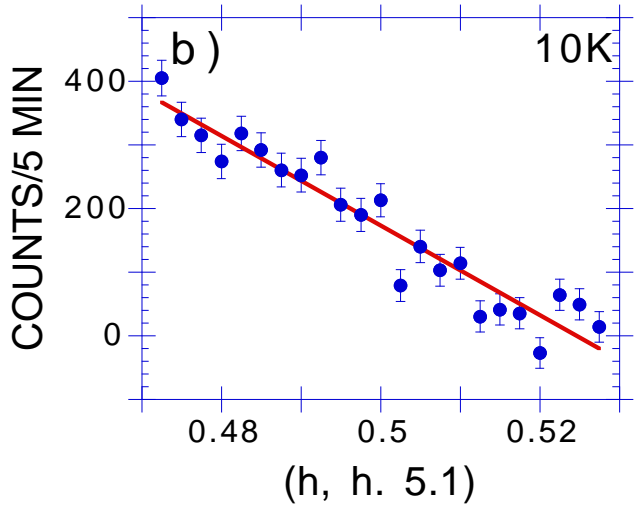
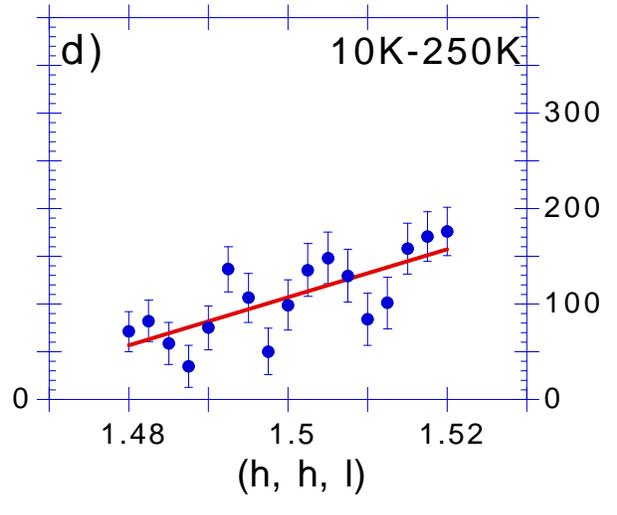
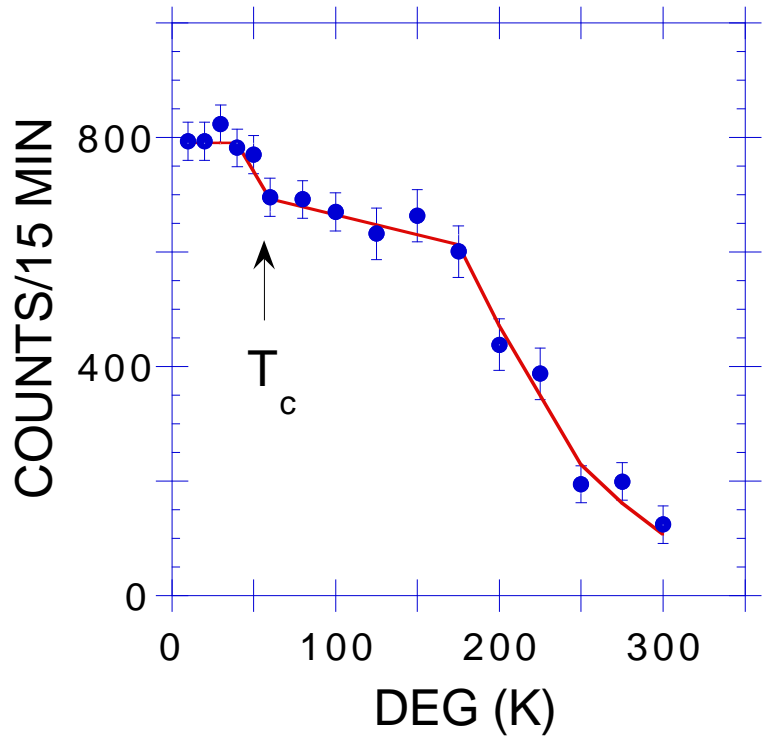
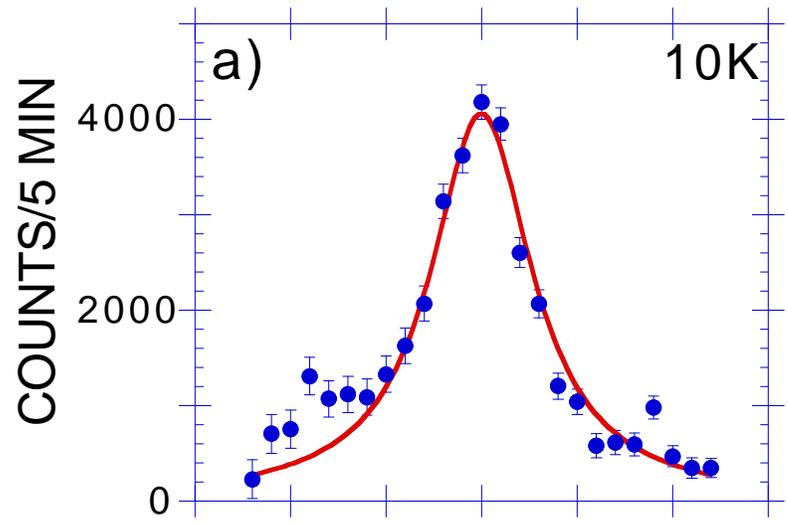
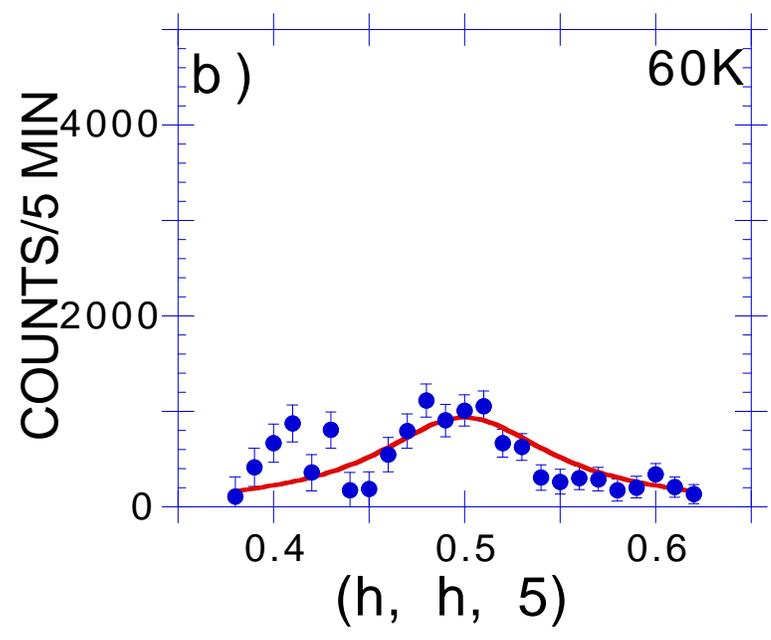
\end{document}